\documentclass[12pt]{iopart}

\usepackage{hyperref}
\usepackage{siunitx}
\usepackage[utf8]{inputenc}
\usepackage{graphicx}
\usepackage{braket}
\usepackage{xcolor}
\graphicspath{{C:/Users/Johannes/Desktop/PaperWritten_StandWarschau/paperdrafts/Bilder/}}
\begin{document}
\bibliographystyle{iopart-num}

\title[]{Spectroscopic Investigations of Negatively Charged Tin-Vacancy Centres in Diamond}

\author{Johannes Görlitz$^1$, Dennis Herrmann$^1$, Gergő Thiering$^2$, Philipp Fuchs$^1$, Morgane Gandil$^1$, Takayuki Iwasaki$^3$, Takashi Taniguchi$^4$, Michael Kieschnick$^5$, Jan Meijer$^5$, Mutsuko Hatano$^3$, Adam Gali$^2$, and Christoph Becher$^1$}

\address{$^1$Fachrichtung Physik, Universität des Saarlandes, Campus E2.6, 66123 Saarbrücken, Germany}
\address{$^2$Wigner Research Centre for Physics, Institute for Solid State Physics and Optics, Hungarian Academy of Sciences, Hungary}
\address{$^3$Department of Electrical and Electronic Engineering, Tokyo Institute of Technology, Meguro, Tokyo 152-8552, Japan}
\address{$^4$Advanced Materials Laboratory, National Institute for Material Science, 1-1 Namiki, Tsukuba, 305-0044, Japan}
\address{$^5$Division of Applied Quantum System, Felix Bloch Institute for Solid State Physics, University of Leipzig, 04103, Leipzig, Germany}
\ead{christoph.becher@physik.uni-saarland.de}
\vspace{10pt}
\begin{indented}
\item[]July 2019
\end{indented}

\begin{abstract}
The recently discovered negatively charged tin-vacancy centre in diamond  is a promising candidate for applications in quantum information processing (QIP). We here present a detailed spectroscopic study encompassing single photon emission and polarisation properties, the temperature dependence of emission spectra as well as a detailed analysis of the phonon sideband and Debye-Waller factor. Using photoluminescence excitation spectroscopy (PLE) we probe an energetically higher lying excited state and prove fully lifetime limited linewidths of single emitters at cryogenic temperatures. For these emitters we also investigate the stability of the charge state under resonant excitation. These results provide a detailed insight into the spectroscopic properties of the $\text{SnV}^-$ centre and lay the foundation for further studies regarding its suitability in QIP.
\end{abstract}

%
%
%
%
%

\section{Introduction}

As the field of QIP is rapidly expanding, significant effort has been spent on the search for quantum bit systems providing favourable properties such as individually addressable spins with long coherence times, optical emission spectra with narrow homogeneous and inhomogeneous broadenings and bright single photon emission. At the same time, it is necessary to maintain scalability of such devices which is an advantage inherent to solid state emitters. Among potential candidates \cite{AtatureMatPlatRev, AwschalomQuTechSSSRev}, colour centres in diamond were shown to be highly suitable to encounter many of the tasks set in QIP such as coherent manipulation of single spins with long coherence times \cite{Bar-GillNVCohTime, AbobeihOneSecNV, BeckerMilliK, SukachevQMSiV}, single photon non-linearities \cite{SipahigilSPSwitch}, strong light-matter interactions \cite{ WeinzetlCohContWaveMix} and entanglement of remote spins \cite{SipahigilQuIntRemNV, HensenLoopholeBell}. More specifically, two colour centres raised strong interest, i.e. the well studied negatively charged nitrogen-vacancy centre ($\text{NV}^-$) due to its spin coherence times approaching one second \cite{AbobeihOneSecNV} and the negatively charged silicon-vacancy centre ($\text{SiV}^-$) with its narrow, bright zero phonon line (ZPL) \cite{NeuSPSiVNano} and the optically addressable, doubly spin degenerate four level fine structure \cite{HeppElecStr, BeckerUltrafastAllOpt}. Nevertheless, both of them suffer from severe drawbacks. In particular the $\text{NV}^-$ centre is subject to strong spectral diffusion \cite{FuSpecDiffNV, WoltersUltrafastSpecDiffNV} and low photon emission into its ZPL, while the $\text{SiV}^-$ centre reaches long spin coherence times only at millikelvin temperatures, requiring the use of a helium dilution refrigerator \cite{BeckerMilliK, SukachevQMSiV}.
\newline 
For this reason there is an ongoing search for a colour centre combining the advantages of both the $\text{NV}^-$ and the $\text{SiV}^-$ centre. A potential candidate for this is the negatively charged tin-vacancy centre ($\text{SnV}^-$ centre) \cite{HatanoTin, TchernijSnV}, exhibiting a four level fine structure, similar to that of the $\text{SiV}^-$ centre. Simulations predict that it shares the same molecular structure, consisting of an impurity atom in a split vacancy configuration in D$_{\text{3d}}$ symmetry, with all group IV - vacancy (G4V) centres (SiV, GeV, SnV, PbV) \cite{ThieringGroupIV}. The inherent inversion symmetry of this structure renders the G4V centres insensitive to first order stark shifts, resulting in small spectral diffusion. On the other hand, the ground state splitting of the $\text{SnV}^-$ centre amounts to 850 GHz, which is sufficient to overcome phonon mediated dephasing processes \cite{JahnkeElecPhonSiV}, that are the main source of decoherence for the $\text{SiV}^-$ centre \cite{PingaultCohDarkStates, RogersCohPrepSiV}, already at liquid helium temperatures \cite{HatanoTin}. This renders the $\text{SnV}^-$ centre an interesting candidate for application in QIP. 
\newline
In this work, we characterise the optical emission properties of single as well as ensembles of negatively charged SnV centres. By investigating the temperature dependence of emission spectra we furthermore prove the suitability of $\text{SnV}^-$ defects as temperature sensors at nanoscale. The polarisation of the defect in emission as well as in absorption is providing evidence for an alignment of the defect in the crystallographic $\langle 111\rangle$ direction of the diamond host matrix. At cryogenic temperatures we prove truly Fourier limited linewidths and investigate the stability of the charge state of single emitters. Additionally the temperature dependence of the Debye-Waller factor and a detailed analysis of the phonon sideband at cryogenic temperatures reveals a dominant coupling of the centre to phonons of $a_{1g}$ symmetry. Finally we probe an energetically higher lying excited state using photoluminescence excitation spectroscopy and thereby extend the understanding of the level structure of the $\text{SnV}^-$.

\section{Investigated samples}\label{samples}

\begin{figure}[h!]
\includegraphics[width=\linewidth]{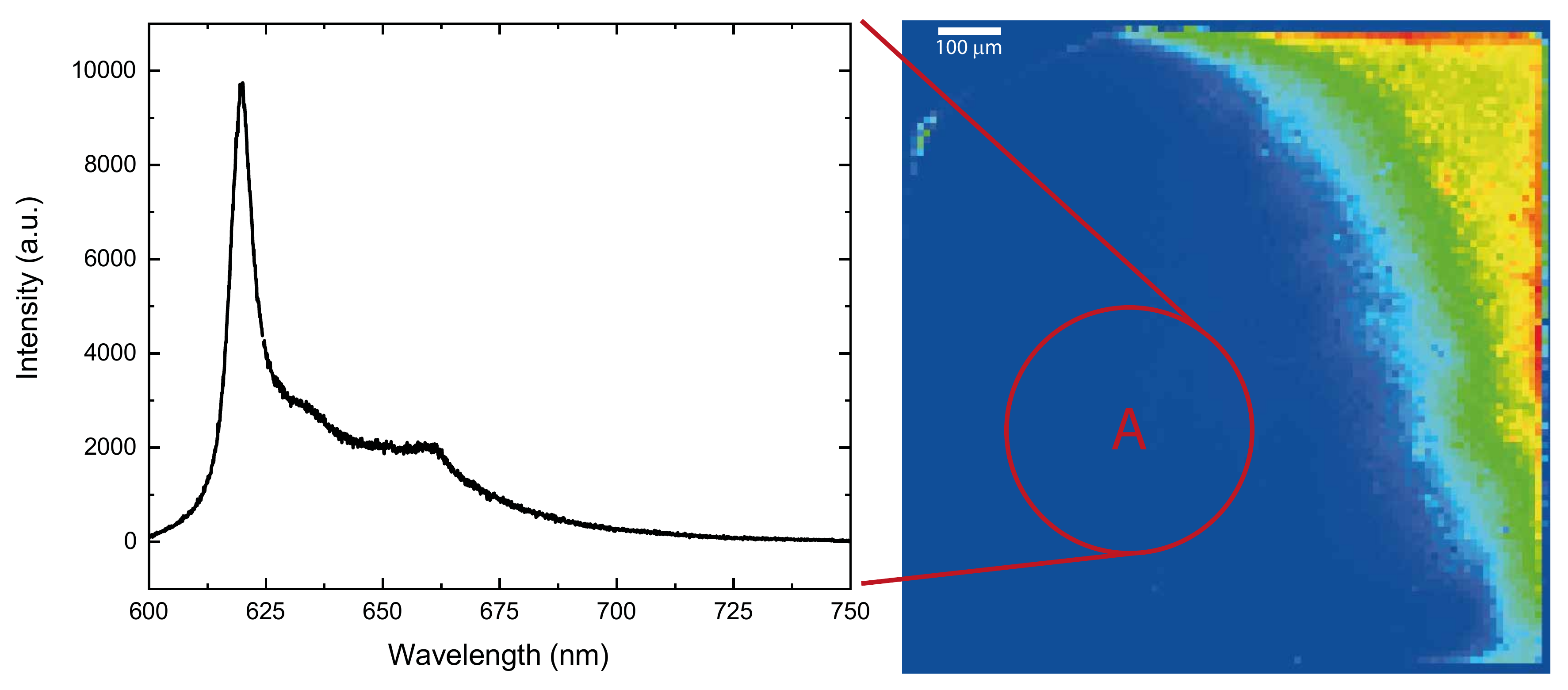}
\caption{Fluorescence scan of sample NI58 under off-resonant \SI{532}{nm} excitation. The crescent shape of the fluorescence pattern is due to inhomogeneously removed diamond during the annealing process. In the bright regions ensembles of varying density of $\text{SnV}^-$ centres are situated, while in section A single $\text{SnV}^-$ centres can be distinguished. An exemplary room temperature spectrum of such a single $\text{SnV}^-$ with the dominant ZPL centered at \SI{620}{nm} and its adjacent phononic sideband ranging up to \SI{750}{nm} is shown.}
\label{fig:Fluo}
\end{figure}

The predominantly investigated sample NI58 in this study is an (001) electronic grade bulk diamond homogeneously implanted with tin ions employing an implantation energy of \SI{700}{keV} and a dose of $2\ \text{x}\ 10^{13}\ \frac{\text{Ions}}{\text{cm}^2}$. The lateral homogeneity of implantation was achieved by scanning the ion beam (diameter of \SI{1}{cm^2}) repeatedly over the sample. Afterwards, an annealing at \SI{2100}{\celsius} with a pressure of \SI{7.7}{GPa} is performed analogous to the procedure reported in \cite{HatanoTin}. In Fig. \ref{fig:Fluo} a fluorescence scan of the diamond sample with \SI{532}{nm} excitation is shown, revealing a crescent-shaped fluorescence pattern of $\text{SnV}^-$ centres. Inhomogeneous removal of the implanted diamond surface during the annealing process provides a smooth transition from very dense ensembles of $\text{SnV}^-$ centres (upper right corner) to regions where single emitters can be found (Region A in Fig. \ref{fig:Fluo}). By this means, we are able to study the spectral properties of ensembles as well as single $\text{SnV}^-$ centres implanted and annealed under the same conditions in one sample. An exemplary room temperature spectrum of such an emitter is also shown in Fig. \ref{fig:Fluo} with its characteristic zero phonon line (ZPL) at \SI{620}{nm}.
For comparison, a second sample SC500\_01 is investigated, for which tin ions were implanted in an (001) electronic grade bulk diamond with an energy of \SI{80}{keV} and varying doses resulting in an mean implantation depth of \SI{26}{nm} (obtained by Monte-Carlo simulations, \textit{Stopping and Range of Ions in Matter}, SRIM). After implantation, the sample is annealed at \SI{1200}{\celsius} for \SI{4}{h} in high vacuum ($10^{-6}$\ mbar). Both samples were cleaned in boiling tri-acid (1:1:1 mixture of sulfuric, perchloric and nitric acid) for \SI{2}{h} at \SI{500}{\celsius}. Sample SC500\_01 was additionally oxidized at \SI{450}{\celsius} in an air atmosphere.
\newline
All samples are investigated in a home built confocal microscope with various excitation sources. Experiments at cryogenic temperatures make use of a liquid helium flow cryostat (Janis Research, ST-500LN) or a closed cycle system (attodry2100, attocube).

\section{Single Photon Emission and Fluorescence Lifetime}

\begin{figure}[h!]
\includegraphics[width=\linewidth]{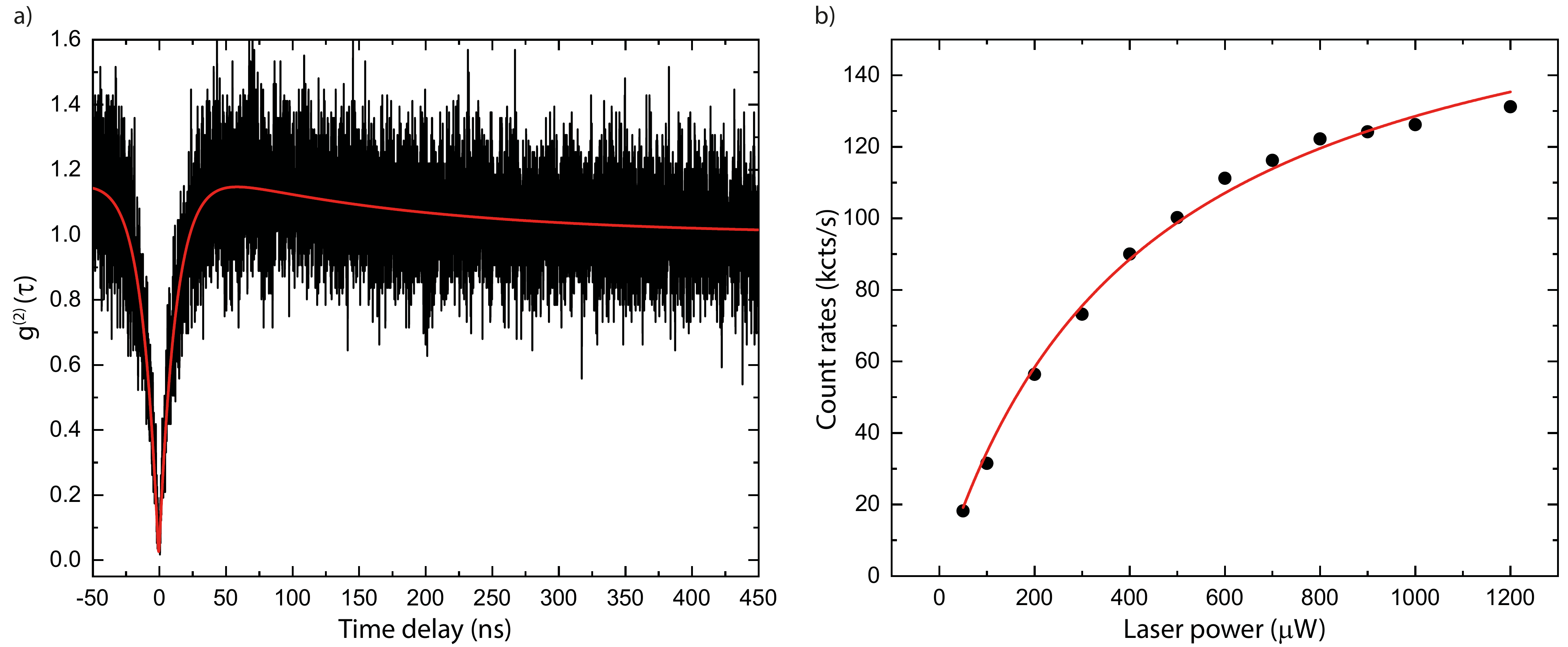}
\caption{a) Raw data of the autocorrelation measurement on a single $\text{SnV}^-$ emitter revealing a $g^{(2)}(0)$ value of 0.05. The deviation from 0 can be fully explained by the dark counts of the APDs. b) Typical saturation measurement of a single $\text{SnV}^-$ emitter. Saturation count rates vary between \SI{80} and \SI{150}{kcts/s} for saturation intensities of 200-\SI{600}{\mu W}. No discernible linear background contribution is visible. The measurements are carried out using off-resonant \SI{532}{nm} excitation.}
\label{fig:g2andSat}
\end{figure}

\begin{figure}[h!]
\includegraphics[width=\linewidth]{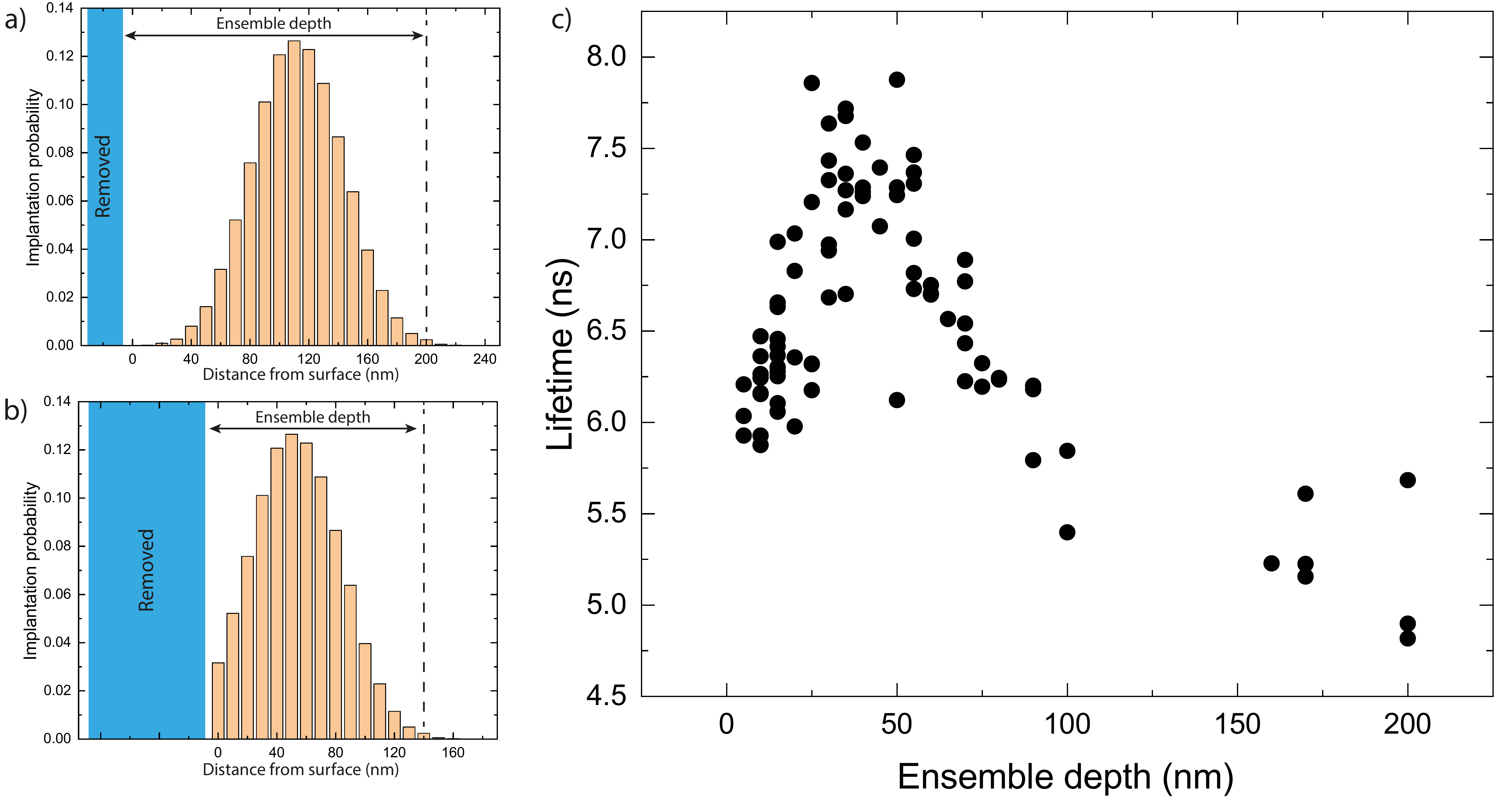}
\caption{The ensemble depth is defined as the distance from the surface (blue shaded area) till the wing of the Gaussian distribution (simulated implantation depth of the tin atoms obtained by using SRIM Monte-Carlo simulations) reaches approximately one percent of its total height. The determination of this depth in the experiment is explained in the main text in detail. Schematically, the ensemble distribution for an ensemble depth of a) 200 and b) \SI{140}{nm} is depicted. In c) the change in lifetime with ensemble depth is shown. The increase in lifetime from 200 to \SI{40}{nm} is caused by reduction of the photonic local density of states, due to increasing number of emitters situated close to the surface. The subsequent drop is most likely caused by non radiative decay channels opening up as it was also seen in \cite{InamTrackEmRate}. The lifetime is measured using off-resonant pulsed excitation with \SI{532}{nm}.}
\label{fig:EnsLifetime}
\end{figure}

In order to prove that our investigations are performed on single $\text{SnV}^-$ centres, we measure the intensity autocorrelation function $g^{(2)}(\tau)$ of each centre in sample NI58. As it can be seen from the $g^{(2)}$ measurement in Fig. \ref{fig:g2andSat} a), we are able to clearly separate single $\text{SnV}^-$ centres. It has to be emphasized that the plotted data is raw data without any background correction applied. Saturation measurements (\ref{fig:g2andSat} b)) reveal a vanishing background emission. The value of $g^{(2)}(0)=0.05$ can be fully explained by the dark counts of the avalanche photo diodes (APD; SPCM-AQRH-14, Excelitas) being used in the experiment and thereby proving perfect single photon emission. The saturation count rates (intensities) of the single emitters investigated in sample NI58 vary between \SI{80} and \SI{150}{kcts/s} (200-\SI{600}{\mu W}) when employing an NA = 0.9 objective, similar to what was found in \cite{HatanoTin,TchernijSnV}. The count rates are one to two orders of magnitude larger than for $\text{SiV}^-$ centres in unstructured electronic grade bulk diamond \cite{BeckerMilliK, SukachevQMSiV}. Furthermore, we compare the saturation count rates to the largest values for $\text{NV}^-$ centres in bulk diamond known to us \cite{LesikPerfPrefOrientNV}. Taking into account the different efficiencies in collecting photons and the Debye-Waller factors of 4\% (60\%) of the $\text{NV}^-$ ($\text{SnV}^-$) centres, respectively, we find that for $\text{SnV}^-$ centres the single photon count rates into the ZPL exceed the corresponding value for $\text{NV}^-$ centres by roughly one order of magnitude, which potentially enables high repetition rates in quantum information processing schemes. 

From the photon correlation measurement in Fig. \ref{fig:g2andSat} a) we  infer a lifetime of the upper fluorescent level of approximately \SI{13}{ns}. To further explore the timescale of the fluorescence decay, we measure the radiative decay lifetime for ensembles of different densities by time correlated single photon counting in sample NI58. The density in this sample is an indicator for the depth of the ensemble within the diamond lattice since diamond was removed inhomogeneously during sample preparation, see section \ref{samples}. We define the depth of an investigated ensemble by the distance from the surface at which the implantation probability (obtained by Monte-Carlo simulations, SRIM) has dropped to approximately one percent of the distribution maximum (which lies at \SI{168}{nm} depth for implantation without removal of diamond). This is illustrated in Fig. \ref{fig:EnsLifetime} a) and b) for the case of 200 and \SI{140}{nm} ensemble depth. The ensemble depth in the experiment is determined by first measuring the count rates in a spot in the high density region where no diamond was removed during sample preparation. This value is used to normalise the area of the Gaussian implantation distribution. For every investigated ensemble we measure the count rates and compare this to our normalised distribution, i.e. until which depth the distribution has to be integrated to obtain the measured count rates. It has to be noted that the count rates are measured at excitation powers far below saturation and therefore the change in lifetime is not affecting the value significantly. In Fig. \ref{fig:EnsLifetime} c) the lifetime of a large set of different ensembles is plotted against the ensemble depth below the diamond surface. Ensembles being deeply situated within the crystal lattice ($>$\SI{100}{nm}) exhibit lifetimes on the order of \SI{5}{ns} as it was theoretically predicted and experimentally observed before in several studies \cite{ HatanoTin,TchernijSnV, ThieringGroupIV, RugarCharSnVPill, TrusheimTransLimSnV}. With decreasing ensemble depth the lifetime increases due to a reduction of the photonic local density of states for large fractions of the ensemble being located in proximity to the surface as it was seen for NV centres before \cite{MohtashamiSuitNVNDSpontEmExp}. At an ensemble depth of about \SI{40}{nm} the lifetime distribution reaches a maximum and decreases steeply when approaching zero distance to the surface, which can be explained by non radiative decay channels opening up for emitters in very close vicinity to the surface. This is in agreement with \cite{InamTrackEmRate} having observed similar variations in lifetime of $\text{NV}^-$ centres with the dimension of the nanodiamonds they are incorporated in. For the investigated single emitters, lifetimes vary between 7 and \SI{25}{ns} which is significantly longer than what was observed before \cite{HatanoTin, TchernijSnV, RugarCharSnVPill, TrusheimTransLimSnV}. In sample NI58 all single emitters are in close proximity to the surface due to the sample fabrication. Removal of diamond during annealing procedure introduces damage on the surface of the diamond and therefore leaves single emitters in a damaged environment close to the surface, which is most likely the source for the large variations in lifetime.

\section{Zero Phonon Line Emission}

\subsection{Temperature Dependent Spectroscopic Investigation of $\text{SnV}^-$ Centres}

\begin{figure}[h!]
  \includegraphics[width=\linewidth]{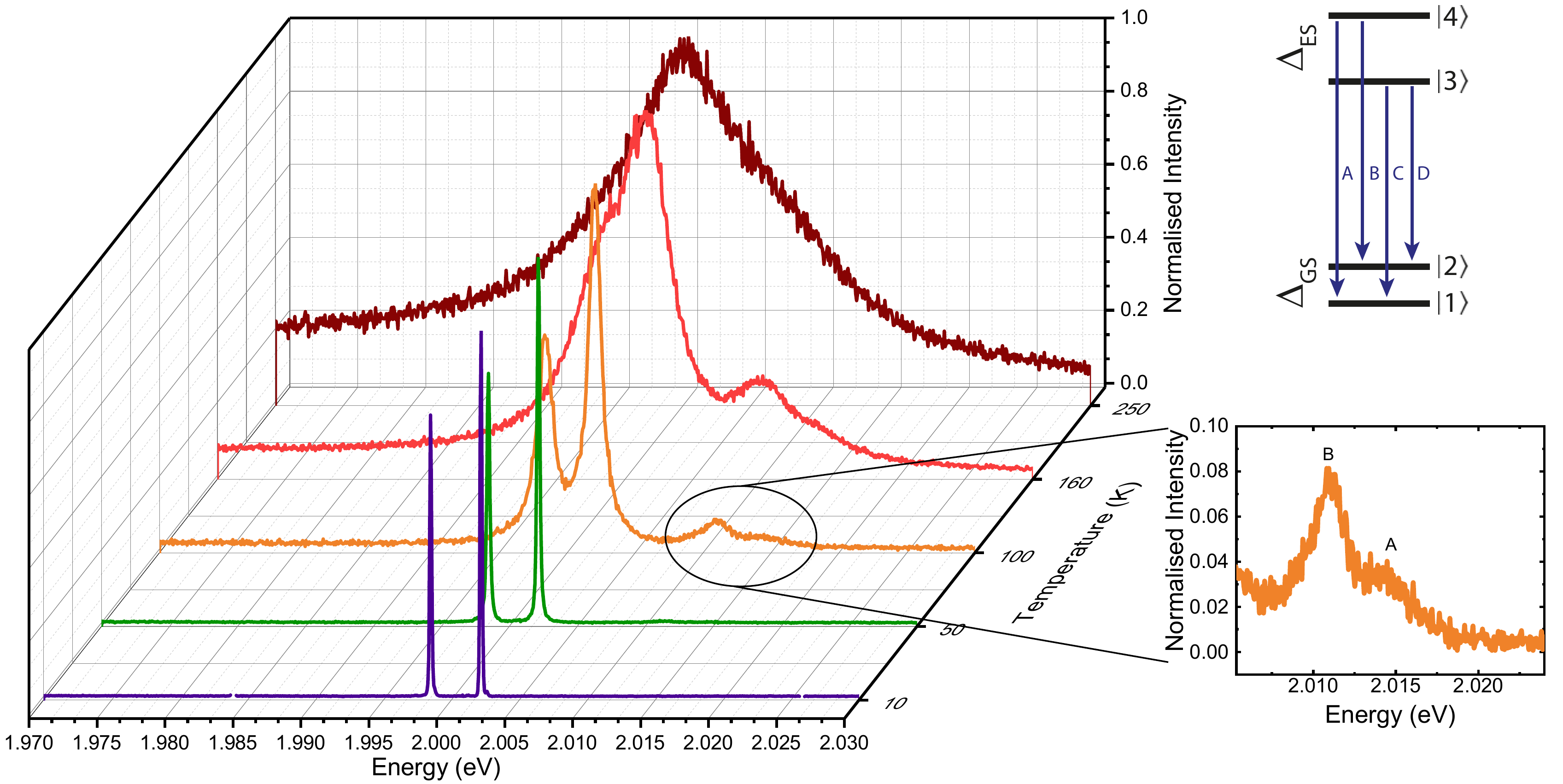}
  \caption{Spectra of a single emitter under off-resonant \SI{532}{nm} excitation at temperatures in between \SI{10}{K} and \SI{250}{K}. At \SI{10}{K} only the C- and D-Transitions are visible, at temperatures around \SI{75}{K} also A and B occur due to increasing thermal population in the upper excited state. With rising temperature, a line shift as well as line broadening is observable.}
  \label{fig:WaterfallS2}
\end{figure}

\begin{figure}[h!]
  \includegraphics[width=\linewidth]{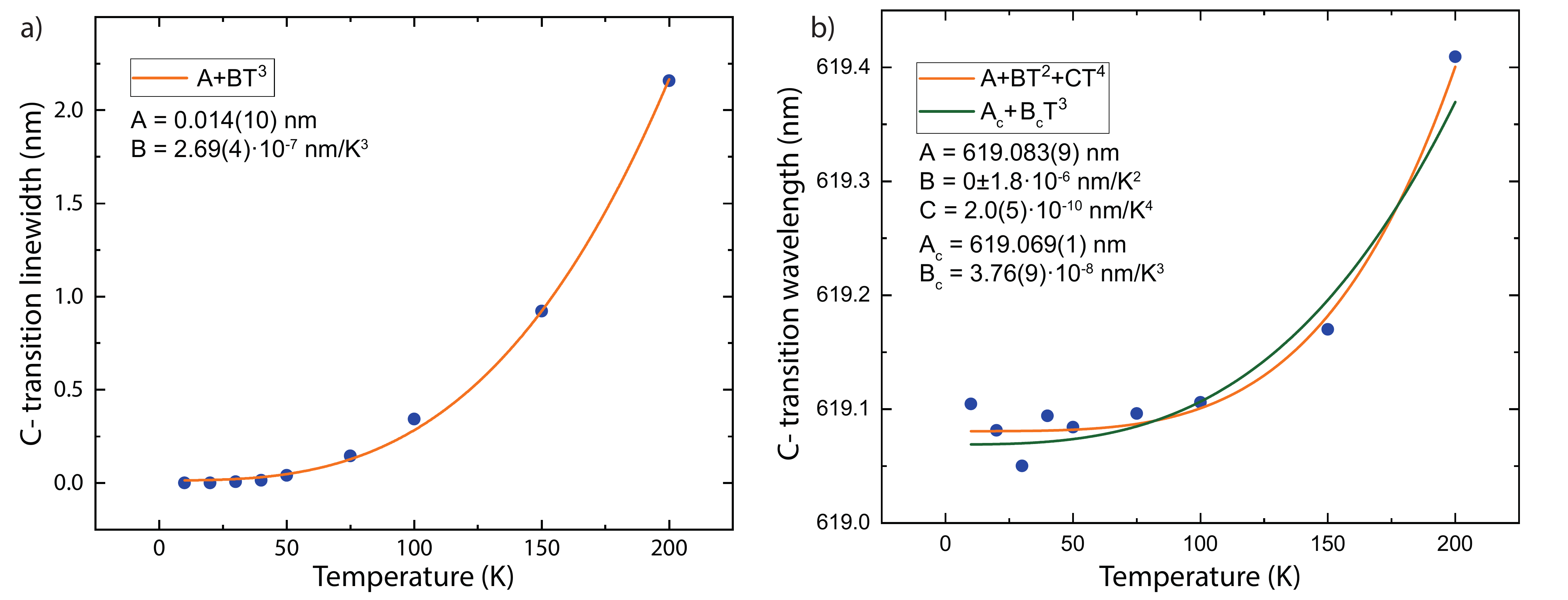}
  \caption{Lineshift and linewidth for increasing temperature are depicted for the C-transition of a single emitter. Dots are experimentally obtained values, the solid lines are fits regarding to the power laws mentioned in the text. The measurements are carried out using off-resonant \SI{532}{nm} excitation.}
  \label{fig:LineWidthShiftE58}
\end{figure}

In order to characterise the electron-phonon interaction of $\text{SnV}^-$ centres, we first measure temperature dependent spectra of single defects in a temperature range from \SI{10}{K} up to \SI{250}{K} in sample NI58. As it can be seen from Fig. \ref{fig:WaterfallS2}, at \SI{10}{K} only two optical transitions C and D occur. These correspond to the decay from the lower excited state $\ket{3}$ into the ground states $\ket{2}$ and $\ket{1}$. With rising temperature the thermal population in the upper excited state $\ket{4}$ increases. The energy separation in the excited state doublet averaged over several measurements yields \SI{3030(100)}{GHz}. Thus transitions A and B only become visible at elevated temperatures as was also seen in \cite{HatanoTin}. The $\text{SnV}^-$ thus exhibits the same fine structure as the $\text{SiV}^-$ for which a $\text{D}_{3\text{d}}$ symmetric defect orientation in the diamond lattice was proven \cite{HeppElecStr}. The increasing linewidth of the transitions follows a $\text{T}^3$ power law (Fig. \ref{fig:LineWidthShiftE58} a)), which was theoretically suggested and experimentally confirmed for $\text{SiV}^-$ centres as well, indicating a low strain crystal environment \cite{JahnkeElecPhonSiV,ArendSpecHole}. The lineshift is very well described by a $\text{T}^2+\text{T}^4$ power law with a vanishing $\text{T}^2$ contribution. The $\text{T}^3$ power law for a ZPL shift according to linear e-symmetric phonon coupling which was derived by Jahnke et al. \cite{JahnkeElecPhonSiV} is not reproducing the data well (see Fig. \ref{fig:LineWidthShiftE58} b)). However, the accurate description of the linewidth broadening with temperature makes the $\text{SnV}^-$ centre a suitable candidate for low temperature sensing on the nanoscale. Its large ground state splitting of \SI{830(30)}{GHz} (lowest value for ensemble as well as for single emitter measurements) is advantageous in comparison to the $\text{SiV}^-$ centre due to the fact that the four transitions merge at comparatively higher temperatures making the fits more reliable at temperatures below \SI{75}{K} and thereby covering a larger temperature range with high resolution.

\begin{figure}[h!]
  \includegraphics[width=\linewidth]{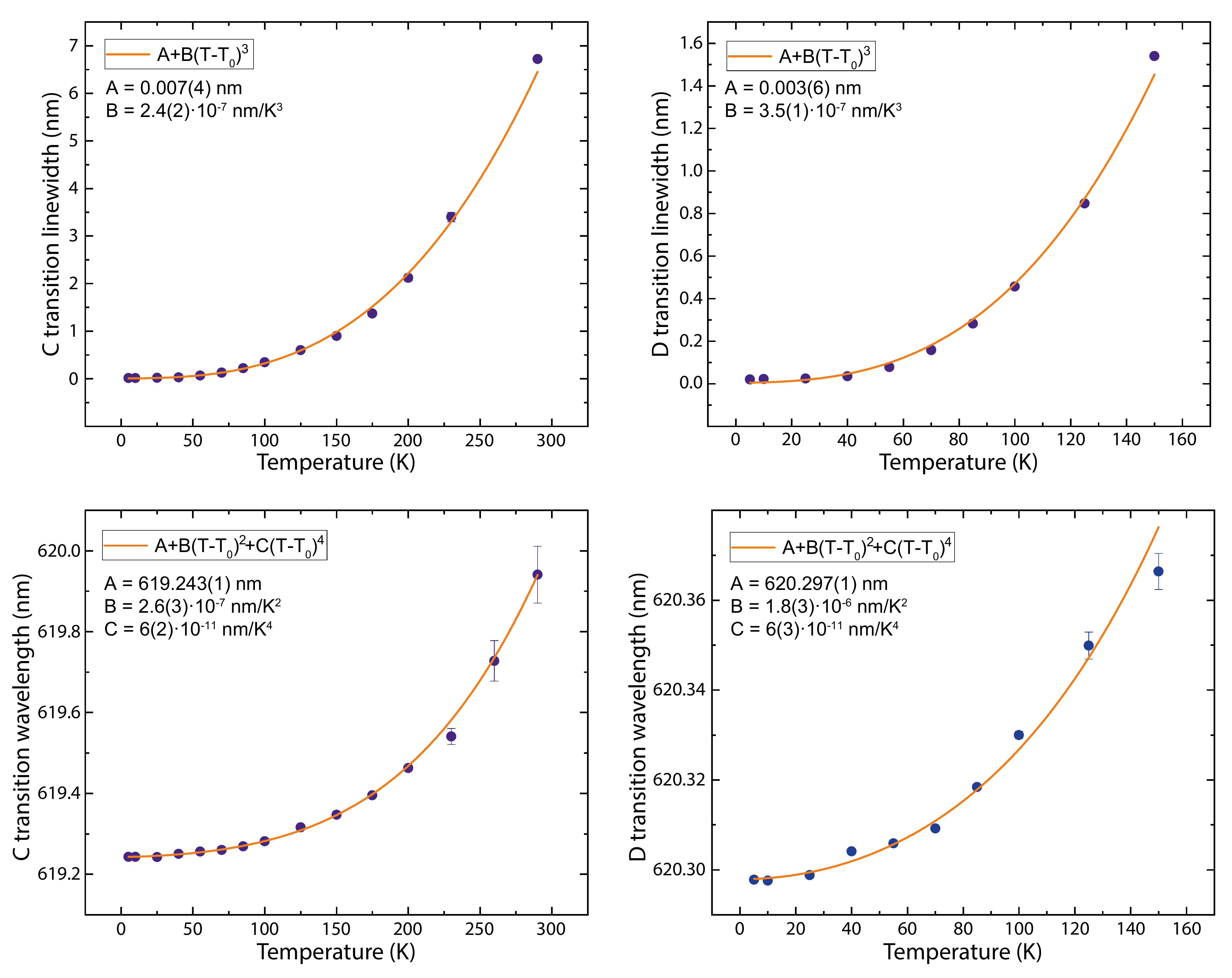}
  \caption{Linewidth and lineshift for increasing temperature are depicted for the C- and D-transition of a spectrally narrow ensemble of $\text{SnV}^-$ centres. Dots refer to measured values while the solid lines are fits regarding to the power laws mentioned in the text. A temperature offset of $\text{T}_0=\SI{10}{K}$ is derived from the fits, most likely caused by improper thermal anchoring after newly mounting the sample in the cryostat. The measurements are carried out using off-resonant \SI{532}{nm} excitation.}
  \label{fig:LineshiftsLinewidthEns}
\end{figure}

The same measurements are repeated for ensembles of $\text{SnV}^-$ centres in sample NI58. To this end, a region on the sample is chosen where the linewidth at a temperature of \SI{5}{K} is below the resolution limit of the spectrometer of \SI{10}{GHz}. Even though the resolution limited linewidth could possibly hint at the investigated number of $\text{SnV}^-$ centres being small, we are not able to see any saturation effects of the ensemble with \SI{10}{mW} of green laser light where count rates exceed \SI{20}{Mcts/s}. For comparison, typical saturation powers for the single emitters investigated in the previous section are on the order of \SI{200}{\mu W}, exhibiting average saturation count rates of \SI{120}{kcts/s}. The $\text{SnV}^-$ ensemble spectrum at cryogenic temperatures consists of the same characteristic finestructure as the single $\text{SnV}^-$ centres.
As it is shown in Fig. \ref{fig:LineshiftsLinewidthEns}, the broadening of the linewidths of peaks C and D is very well reproduced by a pure $\text{T}^3$ law as in the single emitter case. Since adding a $\text{T}^5$ or $\text{T}^7$ term leads to no improvement of the fit, we conclude that the ensemble is situated in a very little strained crystal environment \cite{JahnkeElecPhonSiV}.
Furthermore, also the shift of the optical transitions C and D for the ensemble is shown in Fig. \ref{fig:LineshiftsLinewidthEns}. Both of them obey a $\text{T}^2 + \text{T}^4$ law, again agreeing with the data for the single $\text{SnV}^-$ centre.
As for the single centre, the very accurately described dependence of linewidth and line position on temperature renders ensembles of $\text{SnV}^-$ centers excellent candidates for thermometry on nanoscale. This was proposed by Alkahtani et al. but not investigated below \SI{120}{K} \cite{AlkathaniSnVThermo}.
Furthermore, our investigations hint at very low strain in still dense ensembles of $\text{SnV}^-$ centres which makes them very promising for application in QIP where a high optical density is required while maintaining narrow emission linewidths.

As conclusion of the evaluation of the temperature dependent ZPL emission we find a dominant $\text{T}^4$ contribution for the ZPL line shift of a single $\text{SnV}^-$ centre. An additional $\text{T}^2$ term has to be added to accurately describe the ensemble measurements at temperatures below \SI{100}{K}, which might be associated with inhomogeneous broadening effects. According to Hizhnyakov et al. \cite{HizhnyakovZPLElasticSprings,HizhnyakovTempDepZPLI,HizhnyakovTempDepZPLII} a $\text{T}^4$ law is expected for quadratic electron-phonon coupling of $a_{1g}$ phonons and was also experimentally confirmed for $\text{SiV}^-$ centres \cite{ArendSpecHole}. In Sec. \ref{PSBDW} we present further evidence for the electron-phonon coupling of the optical transition being dominated by $a_{1g}$ phonons.
\newline
The broadening of the ZPL line is caused by single phonon absorption and Raman scattering between sublevels of ground and excited states, respectively, for which a $\text{T}+\text{T}^3$ law has been derived \cite{JahnkeElecPhonSiV}. The sublevels split due to spin-orbit interaction and the corresponding doubly degenerate orbitals are also subject to Jahn-Teller distortion in which the $e_{g}$ phonons will only couple these degenerate orbitals both in the electronic ground and excited state \cite{ThieringGroupIV}. Thus, $a_{1g}$ phonons do not contribute to the ZPL broadening and there only remains the effect of $e_{g}$ phonons. This is in accordance with our observations where the linear contribution $\propto \text{T}$ would only be visible for high resolution measurements at very low temperatures.

\subsection{Comparison of Photoluminescence Spectra for Different Annealing Temperatures}

\begin{figure}[h!]
  \includegraphics[width=\linewidth]{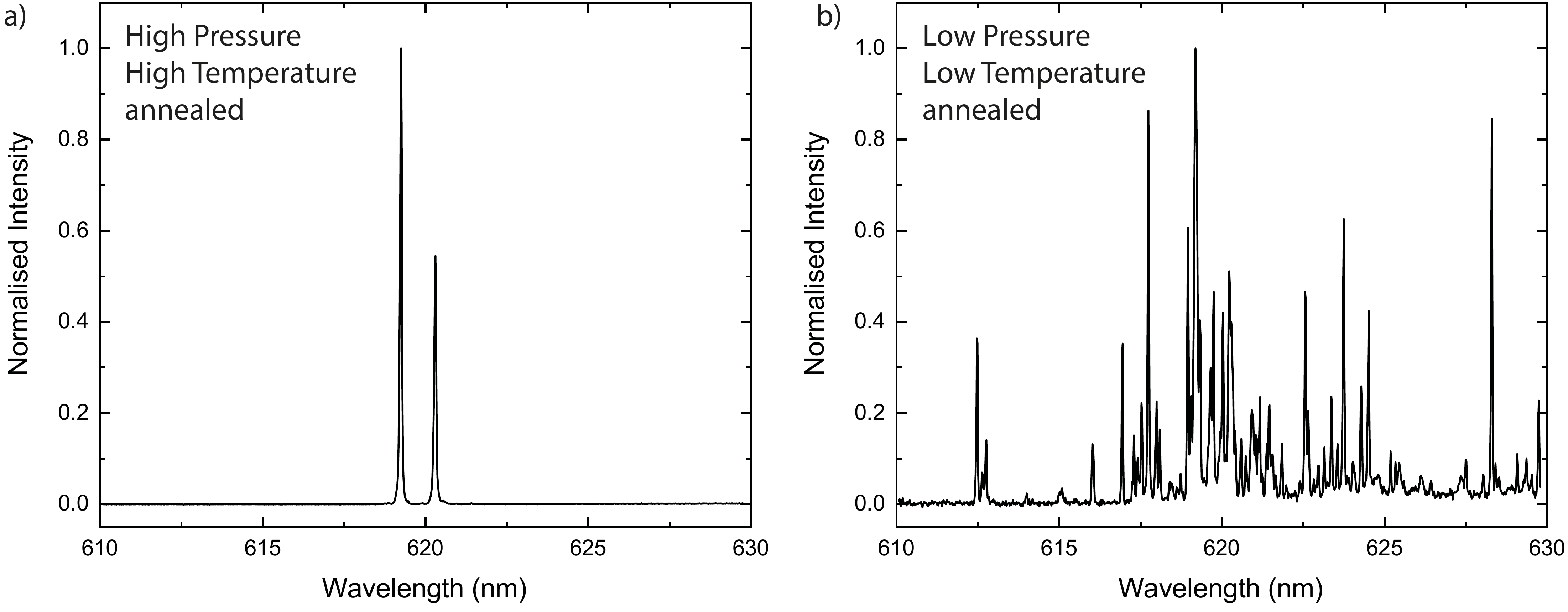}
  \caption{Comparison of ensemble spectra under off-resonant \SI{532}{nm} excitation with similar density of a) sample NI58 annealed at \SI{2100}{\celsius} and \SI{7.7}{GPa} and b) sample SC500\_01 annealed at \SI{1200}{\celsius} and $10^{-6}$\ mbar. The inhomogeneous distribution is tremendously reduced in the higher annealed sample due to removing lattice damages induced by the implantation process. }
  \label{fig:CompEnsSpec}
\end{figure}

\begin{figure}[h!]
  \includegraphics[width=\linewidth]{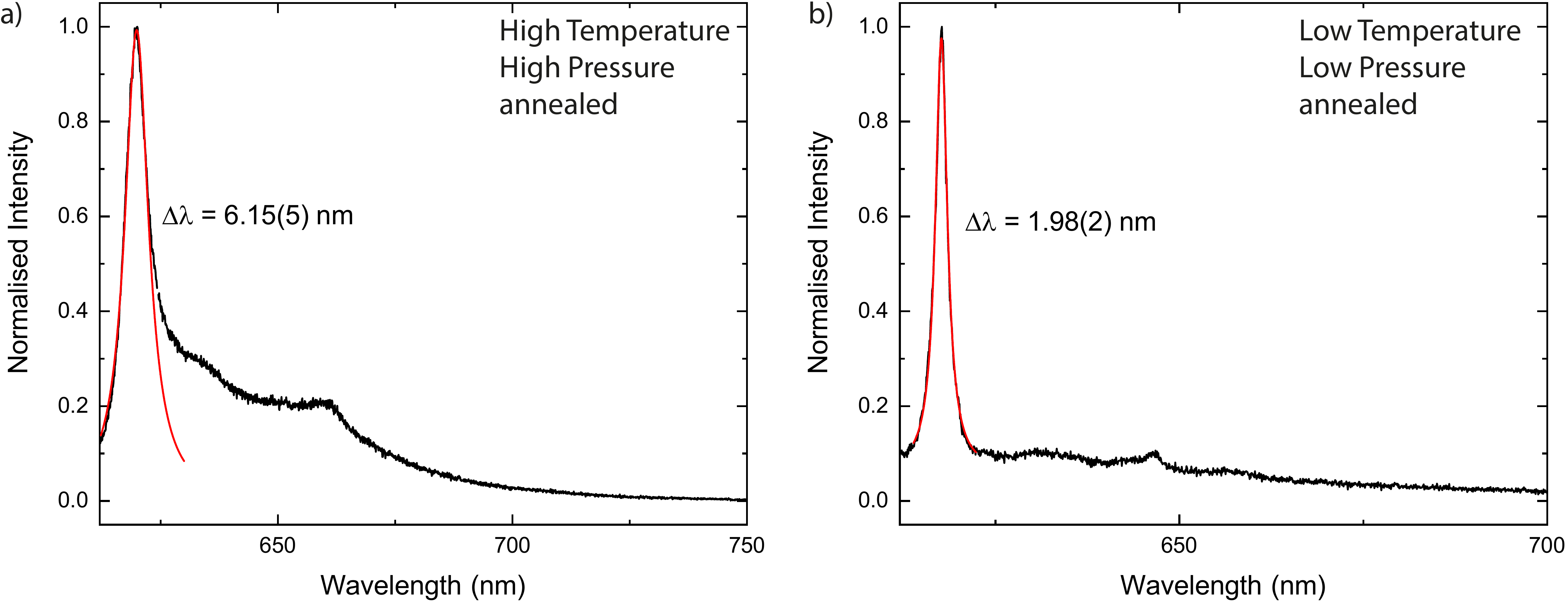}
  \caption{Comparison of typical single $\text{SnV}^-$ room temperature spectra under off-resonant \SI{532}{nm} excitation in a) sample NI58 annealed at \SI{2100}{\celsius} and \SI{7.7}{GPa} and b) sample SC500\_01 annealed at \SI{1200}{\celsius} and $10^{-6}$\ mbar. The ZPL linewidth is significantly smaller in the low-temperature-low-pressure annealed sample than in the high-temperature-high-pressure annealed one.  }
  \label{fig:CompSingleSpec}
\end{figure}

The effect of the annealing temperature after ion implantation on the room temperature linewidth is addressed in \cite{HatanoTin} showing a monotonous decrease of room temperature ZPL linewidth with annealing temperature. We here find that the annealing temperature furthermore very strongly influences the inhomogeneous distribution of ZPL lines. In Fig. \ref{fig:CompEnsSpec}, a comparison between the two investigated samples is shown for ensembles of comparable density of $\text{SnV}^-$ centres. As obvious from Fig. \ref{fig:CompEnsSpec}, the annealing procedure at \SI{2100}{\celsius} greatly reduces the inhomogeneous distribution induced by the lattice damage that is caused during implantation of the heavy tin atoms. The linewidth of the ZPL transitions of the whole ensemble in sample NI58 (Fig. \ref{fig:CompEnsSpec} a)) is \SI{15}{GHz}, which is similar for the individual emission lines (10-\SI{50}{GHz}) occuring in sample SC500\_01. The line positions of the C and D transitions in sample NI58 are within a \SI{150}{GHz} error margin the same for single emitters and ensemble, while they are spread over a large range ($>$\SI{10}{nm}) in sample SC500\_01. We want to emphasize that contrary to the $\text{SiV}^-$ centre, where annealing temperatures of \SI{1200}{\celsius} are sufficient to produce low strain ensembles \cite{ArendSpecHole}, for the $\text{SnV}^-$ centre higher annealing temperatures are required during sample preparation in order to reduce inhomogeneous broadening.
The influence of annealing temperature on strain is also visible for room temperature emission. In Fig. \ref{fig:CompSingleSpec} typical room temperature spectra of single centres are compared for both samples. The ZPL features considerably narrower linewidth (2-\SI{3}{nm}) for the low-pressure-low-temperature (LPLT) annealed sample (Fig. \ref{fig:CompSingleSpec} a)) than for the HPHT annealed NI58 (5-\SI{6.5}{nm}, Fig. \ref{fig:CompSingleSpec} b)). Furthermore, for the HPHT annealed sample we find a very homogeneous distribution of ZPL centre wavelengths around \SI{619.6}{nm} whereas the LPLT annealed sample shows a broad distribution of ZPL centre wavelengths (610-\SI{630}{nm}). In the latter sample we also find reduced emission into the phonon sideband. A reason for this could be a different phonon coupling for different types and strengths of strain in the two samples, resulting in differing phonon broadening. Similar room temperature distributions of ZPL centre wavelengths and linewidths were reported for $\text{SiV}^-$ centres in nanodiamonds, were it is common that large strain is exerted on the colour centres \cite{NeuSPSiVNano,LindnerStrInhSiV}.

\subsection{Polarisation of Single $\text{SnV}^-$ Centres}
\label{sec:Pol}

\begin{figure}[h!]
  \includegraphics[width=\linewidth]{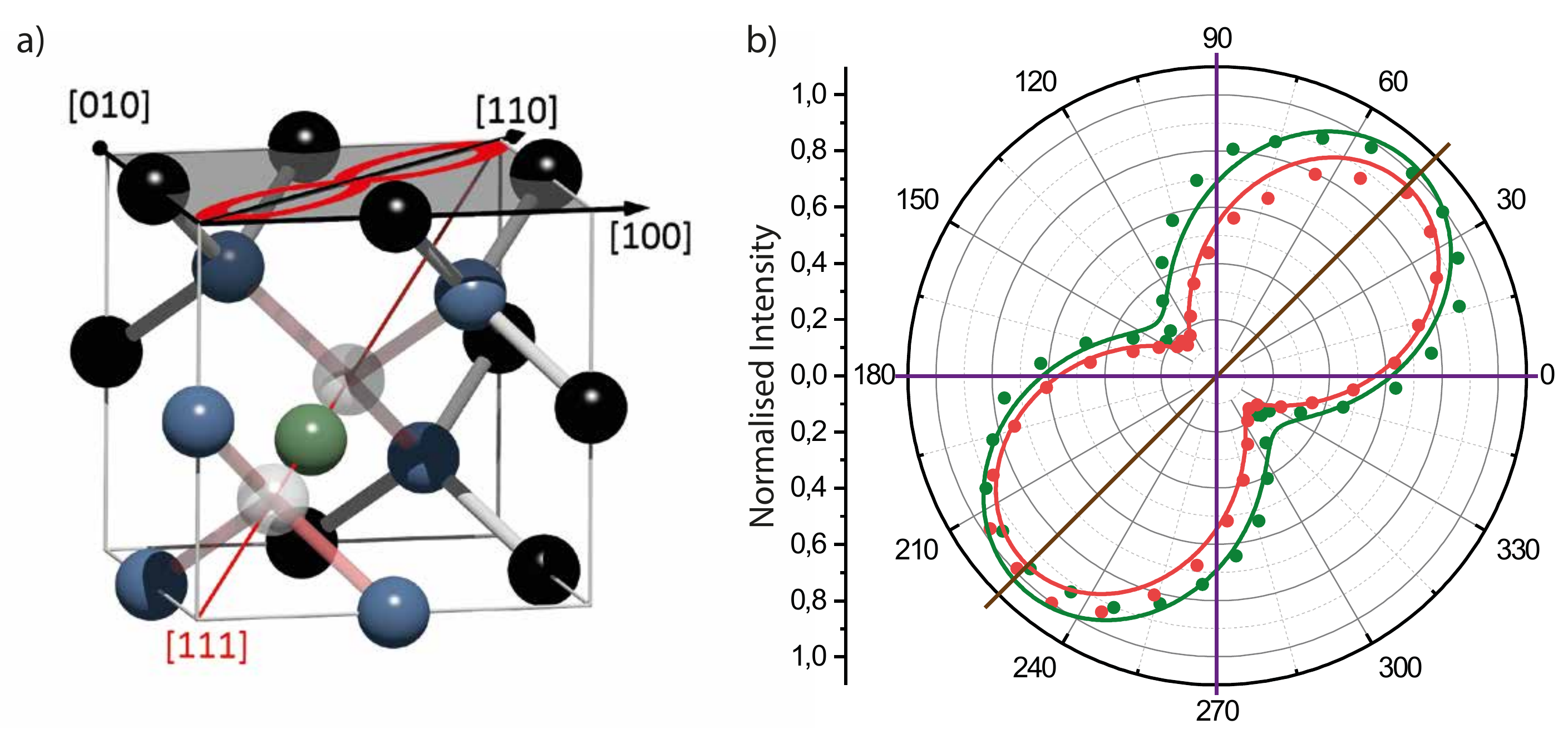}
  \caption{a) Schematic visualisation of the measured dipole projected into the measurement plane (001) b) Absorption and emission dipole measured for a single $\text{SnV}^-$ centre under off-resonant \SI{532}{nm} excitation. Visibilities of up to 85\% in emission and 87\% in excitation are measured for the investigated set of $\text{SnV}^-$ centres. The purple axes indicate the [100] and [010] directions of the diamond lattice. It follows that the main dipole and most probably the high symmetry axis of the $\text{SnV}^-$ center is aligned in $\langle 111\rangle$ direction.}
  \label{fig:PolE57}
\end{figure}

An interesting figure of merit is the polarisation of single $\text{SnV}^-$ centres, which gives insight into the orientation of the dipole within the diamond lattice with respect to the crystal axes. The polarisation of the emitters in sample NI58 is measured in excitation as well as in emission. This is done by employing a linear polariser followed by a half-wave, a quarter-wave plate and finally a dichroic mirror in the excitation pathway. The quarter-wave plate remains at a fixed position throughout the measurements and is used to compensate the phase shift induced by the dichroic mirror. Rotating the half-wave plate and at the same time recording the emitted count rates, reveals the excitation dipole pattern of the single $\text{SnV}^-$ centres. In order to also characterise the polarisation of the emitted photons, a reversed configuration is applied, where the dichroic mirror is followed by a quarter-wave, a half-wave plate and, finally, a linear polariser. Both absorption and emission dipoles are shown for one exemplary $\text{SnV}^-$ centre in Fig. \ref{fig:PolE57} where the count rates are only corrected for the dark counts of the APD. As opposed to the results of \cite{TchernijSnV} it can be seen that the projections of the dipoles into our measurement plane (diamond (001) plane) perfectly overlap and the orientation is at an angle of \SI{45}{\degree} with respect to the surface edges of the sample cut along the [100] and [010] direction. This points towards an alignment of the high symmetry axis of the $\text{SnV}^-$ in $\langle 111\rangle$ direction which is in excellent agreement with former findings for the $\text{SiV}^-$ centre \cite{HeppElecStr, NeuFluoPol}. All measured $\text{SnV}^-$ centres exhibit the same or a perpendicular emission pattern with overlapping absorption and emission dipoles. Visibilities of up to 85\% in emission and 87\% in excitation are measured for the investigated set of $\text{SnV}^-$ centres. The deviation from full contrast is most probably caused by a contribution of weaker x and y dipoles \cite{HeppElecStr}. This is further emphasised by the results of \cite{RugarCharSnVPill} where they find the C-transition being slightly and the D-transition strongly elliptically polarised.

\subsection{Photoluminescence Excitation Spectroscopy and Stability of Tin-Vacancy Centres Under Resonant Excitation}

\begin{figure}[h!]
\includegraphics[width=\linewidth]{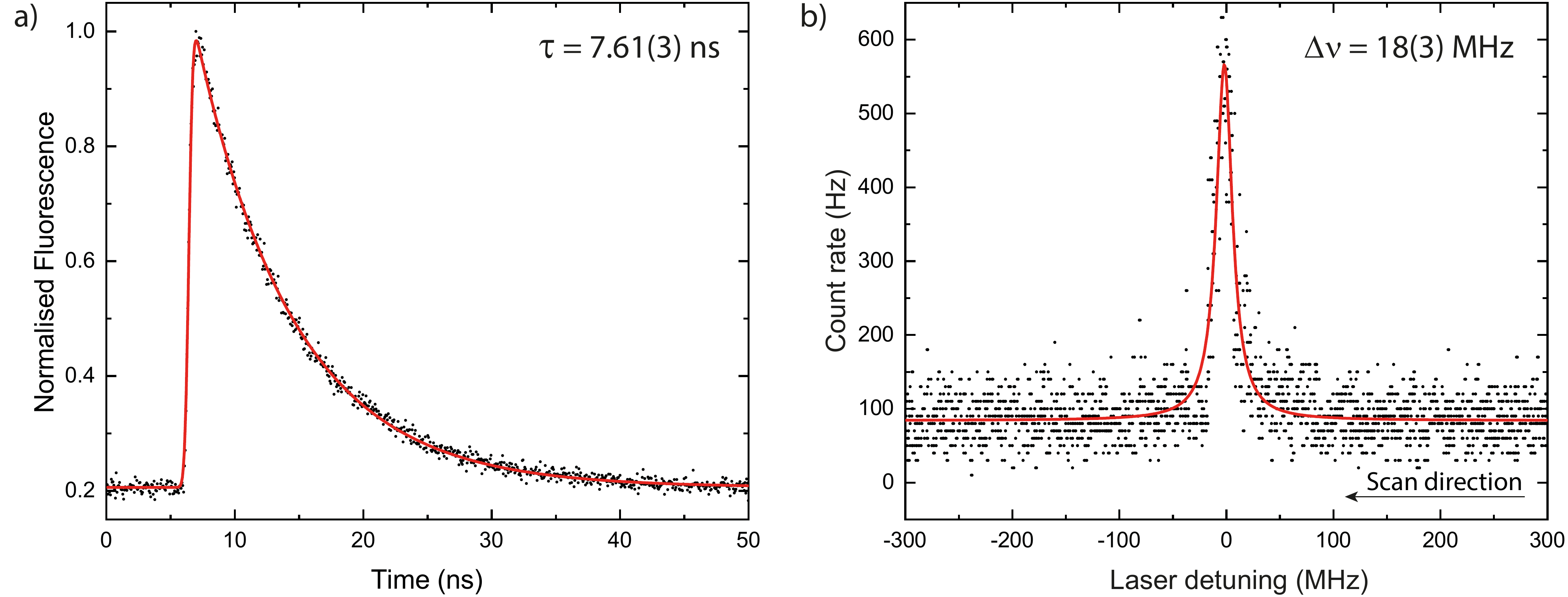}
\caption{a) Fluorescence decay of a single $\text{SnV}^-$ centre under non-resonant excitation with a time constant of \SI{7.61(3)}{ns}. b) PLE spectrum of the C-transition under \SI{200}{pW} excitation power of the same centre exhibiting a truly Fourier limited linewidth of \SI{18(3)}{MHz}.  }
\label{fig:PLE}
\end{figure}

\begin{figure}[h!]
\includegraphics[width=\linewidth]{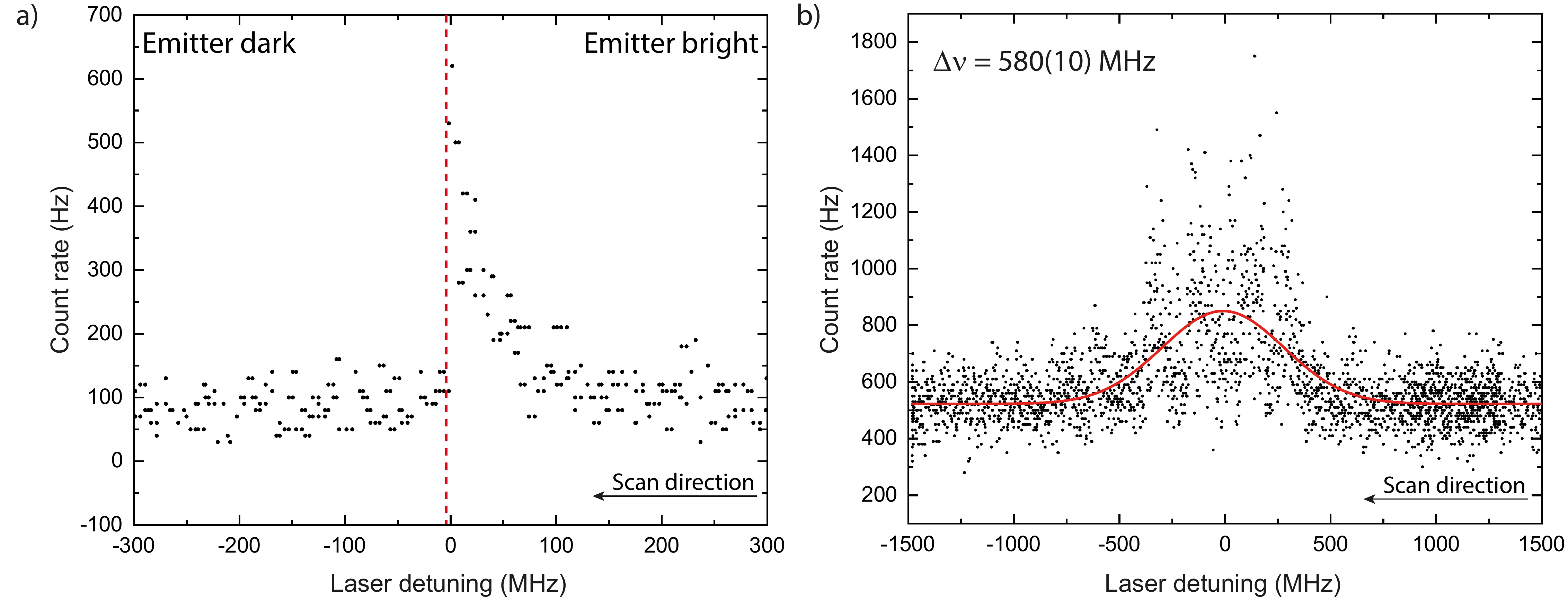}
\caption{PLE measurement of the same single $\text{SnV}^-$ centre as in Fig. \ref{fig:PLE}. a) The emitter is ionized while the laser is scanned over the C-transition. This is visible by the sharp drop of the fluorescence when approaching the resonance maximum. b) An additional \SI{532}{nm} laser with \SI{50}{\micro W} excitation power is continuously added. The linewidth of the C-transition is broadened to a value of \SI{580(10)}{MHz} due to spectral diffusion induced by the dynamic charge equilibrium created by the green laser.}
\label{fig:PLE_Dying_Green}
\end{figure}

The spectral quality of the $\text{SnV}^-$ emission is further determined by the optical coherence of the ZPL at low temperatures. To explore this, we perform photoluminescence excitation (PLE) spectroscopy on the C-transition of single emitters in sample NI58 using excitation with a narrowband Dye laser (Matisse DS, SIRAH) and detecting fluorescence from the excited state into the PSB (filtered in a window from \SI{655}{nm} to \SI{665}{nm}). As it can be seen in Fig. \ref{fig:PLE}, we are able to find purely Fourier-limited linewidths for single $\text{SnV}^-$ centres similar to the findings in \cite{TrusheimTransLimSnV} and significantly narrower than in \cite{RugarCharSnVPill}. For 12 evaluated emission lines we measure a mean linewidth of \SI{31}{MHz}, slightly above the Fourier limit due to power broadening. For the emitter in Fig. \ref{fig:PLE}, the fluorescence lifetime is determined using time correlated single photon counting to be \SI{7.61(3)}{ns} yielding a Fourier limited linewidth of \SI{20.9(1)}{MHz}, being in excellent agreement with the measured value of \SI{18(3)}{MHz}. The excitation power for the measurement in Fig. \ref{fig:PLE} b) is \SI{200}{pW}. We find that even slightly larger excitation powers below \SI{1}{nW} cause termination of the emission, presumably by ionisation from the excited state. Such a termination is exemplarily shown in Fig. \ref{fig:PLE_Dying_Green} a) where a steep drop of fluorescence is visible when approaching the line centre. The charge can be brought back by applying an off-resonant green laser (\SI{532}{nm}, \SI{40}{\mu W}). This is achieved experimentally by coupling the green and resonant lasers into the same fibre to achieve perfect spatial overlap in our confocal microscope and thereby address the same emitter with both colours. We explain this phenomenon with a two-photon ionisation process of $\text{SnV}^-$ to $\text{SnV}^0$ in which the first photon excites the $\text{SnV}^-$ centre resonantly, while the second photon transfers the electron from the excited state to the conduction band. This is in agreement with the ionisation occurring only when approaching the resonance maximum (see Fig. \ref{fig:PLE_Dying_Green} a)) and the calculated (-$\vert$0) ionisation energy of \SI{3.2}{eV} in \cite{ThieringGroupIV} which is exceeded by two \SI{620}{nm} (\SI{2.0}{eV}) photons. Application of the \SI{532}{nm} (\SI{2.33}{eV}) laser leads to a conversion of $\text{SnV}^0$ back to $\text{SnV}^-$ in a single photon process matching again the (0$\vert$-) ionisation energy of \SI{2.3}{eV} derived in \cite{ThieringGroupIV}.
However, the transition line is shifted in this process. Most probably this shift is related  to the second order sensitivity of the $\text{SnV}^-$ centre to an altered charge environment. Constantly adding a green laser while scanning the resonance supports this assumption, as the transition is broadened by spectral jumps to \SI{580(10)}{MHz} as the green laser is creating a dynamic equilibrium of moving charges (see Fig. \ref{fig:PLE_Dying_Green} b)). Similar effects of spectral diffusion were observed for the $\text{SiV}^-$ centre in \cite{NguyenNanophotQuRegSiV}.
\newline 
Since the charge instability could be caused by the emitters being located very close to the surface of the diamond sample, we  further investigate the second sample SC500\_01. The emitters in this sample are situated more than \SI{20}{nm} deep in the diamond lattice but still show the same instabilities under resonant excitation and spectral diffusion occurs in the same manner. We therefore conclude that the negative charge state of the SnV centre is not stable under resonant excitation in an electronic grade diamond host matrix and manipulation of the Fermi level, e.g. by co-implantation of electron donors might be required in order to overcome this problem. 

\section{Analysis of the Phonon Sideband and Debye-Waller Factor}\label{PSBDW}

\begin{figure}[h!]
  \includegraphics[width=\linewidth]{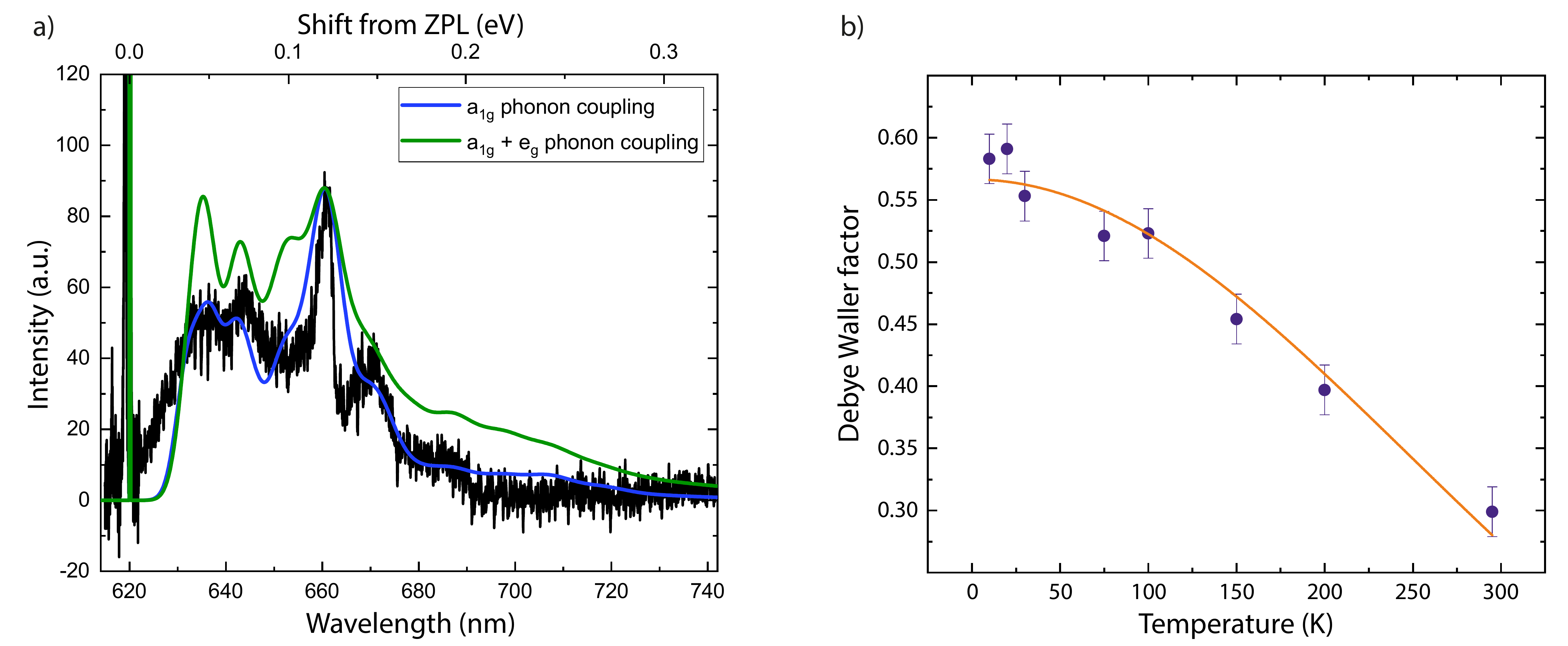}
  \caption{a) Phonon sideband of a single $\text{SnV}^-$ centre at T=\SI{30}{K}. The observable peaks are in excellent agreement with the theoretically predicted phonon sideband for purely $a_{1g}$ phonons coupling to the centre in \cite{ThieringGroupIV} (solid blue line). The combination of $a_{1g}$ and $e_g$ phonon coupling (solid green line) also predicts the peak positions well, but the peak height is less well matching. b) Debye-Waller factor measured for the same centre within a temperature range from \SI{5}{K} up to \SI{300}{K}. The solid line is a fit with the cutoff temperature of the phonons coupling to the centre and the Huang-Rhys factor as free parameters. The measurements are carried out using off-resonant \SI{532}{nm} excitation.}
  \label{fig:PSB_DW}
\end{figure}

Being a solid state emitter embedded in a diamond host matrix renders the $\text{SnV}^-$ centre sensitive to coupling to phononic modes. Since these electron-phonon interactions are of great importance for the spectral purity of the defect emission, a thorough evaluation of the phononic sideband of single $\text{SnV}^-$ centres is necessary. In Fig. \ref{fig:PSB_DW} a) the sideband spectrum of a single $\text{SnV}^-$ in sample NI58 at a temperature of \SI{30}{K} is depicted. There are six observable peaks shifted by \SI{46}{meV}, \SI{76}{meV}, \SI{109}{meV}, \SI{122}{meV}, \SI{148}{meV}, \SI{181}{meV} with respect to the ZPL position. The structure of these peaks is in excellent agreement with the predicted phonon sideband by Thiering and Gali (\cite{ThieringGroupIV}), when only taking into account $a_{1g}$ symmetric phonons coupling to the centre (solid blue line in Fig. \ref{fig:PSB_DW} a)). If comparing merely the position of the peaks, the spectrum for the combined coupling of $a_{1g}$ and $e_g$ phonons (solid green line in Fig. \ref{fig:PSB_DW} a)) also fits, but the peak height is less well predicted. The simulations are normalised to the maximum of the dominant sideband peak at \SI{660}{nm}. Since also the Huang-Rhys factor is calculated for both phonon coupling schemes, the next section is dedicated to a thorough evaluation thereof. It has to be pointed out further that no quasi-local mode caused by the eigenmotion of the tin atom, which should occur at a frequency shift of \SI{29}{meV} from the ZPL (see \cite{TchernijSnV}), can be observed experimentally. This is a further hint at the high quality of the sample since the local oscillator transition corresponding to a coupling to $e_u$ phonons should be symmetry forbidden in an unstrained crystal environment \cite{LonderoVibModesSiV} which in the investigated sample was created by the high temperature annealing. The local mode was observed for $\text{SiV}^-$ and $\text{GeV}^-$ centres \cite{DietrichIsoShiftSiV, EkimovIsoShiftGeV} in material with residual strain. 

The Debye-Waller factor is an important number when evaluating colour centres with respect to their suitability in QIP, as for most of the protocols only photons emitted via the ZPL transition can be used. Therefore, the success probability of such QIP experiments depends directly on the number of photons emitted without phonon interactions. For this reason, we here evaluate the Debye-Waller factor within a temperature range from \SI{5}{K} up to \SI{300}{K} in sample NI58 by fitting the obtained spectra in the range from 610 to \SI{740}{nm} with Voigt profiles in order to account for the Gaussian spectrometer response function. Then the area of the ZPL fit is divided by the total area of the fit to the whole spectrum yielding the Debye-Waller factor. The result is shown in Fig. \ref{fig:PSB_DW} b). The solid line is a fit to the data according to the equation $\text{DW}(\text{T})=\exp(-S(1+\frac{2\pi^{2}}{3}\frac{\text{T}^{2}}{\text{T}_{\text{cutoff}}^{2}}))$ derived in \cite{brand} with S being the Huang-Rhys factor and $\text{T}_{\text{cutoff}}$ the cutoff temperature for the phonons actually coupling to the $\text{SnV}^-$ centre. This cutoff temperature is common to the Debye temperature, as found in \cite{Plakhotnik_OptThermNV}. However, it is not taking into account all phonons present in the diamond lattice but only those involved in the centre-phonon interaction. The fit yields a Huang-Rhys factor of 0.57(2) corresponding to a Debye-Waller factor of DW(\SI{0}{K})=0.57(1) being in close agreement with the theoretically proposed $\text{DW}_{a_{1g}}(\SI{0}{K})=0.63$ for only $a_{1g}$ phonons coupling to the centre \cite{ThieringGroupIV}. The Debye-Waller value for the combined coupling of $a_{1g}$ and $e_g$ symmetric phonons yields $\text{DW}_{a_{1g}+e_g}(\SI{0}{K})=0.41$, again agreeing to a lesser extend with our experiment. The cutoff temperature finally amounts to \SI{680(40)}{K}, which corresponds to an effective phonon frequency of \SI{59(4)}{meV}. The frequency of the acoustical $a_{1g}$ phonons considered in \cite{ThieringGroupIV} was only estimated in a range between 60-\SI{100}{meV} which includes \SI{59(4)}{meV}, while the $e_g$ phonon frequency was calculated to \SI{75.6}{meV}, far off from our measured value and therefore again pointing towards a predominant impact of $a_{1g}$ phonons to the centre-phonon interaction. With this third indicator we can conclude that the lattice coupling between the $\text{SnV}^-$ centre and its host matrix is dominated by phonons of $a_{1g}$ symmetry.

\section{Excited State Spectroscopy}

\begin{figure}[h!]
  \includegraphics[width=\linewidth]{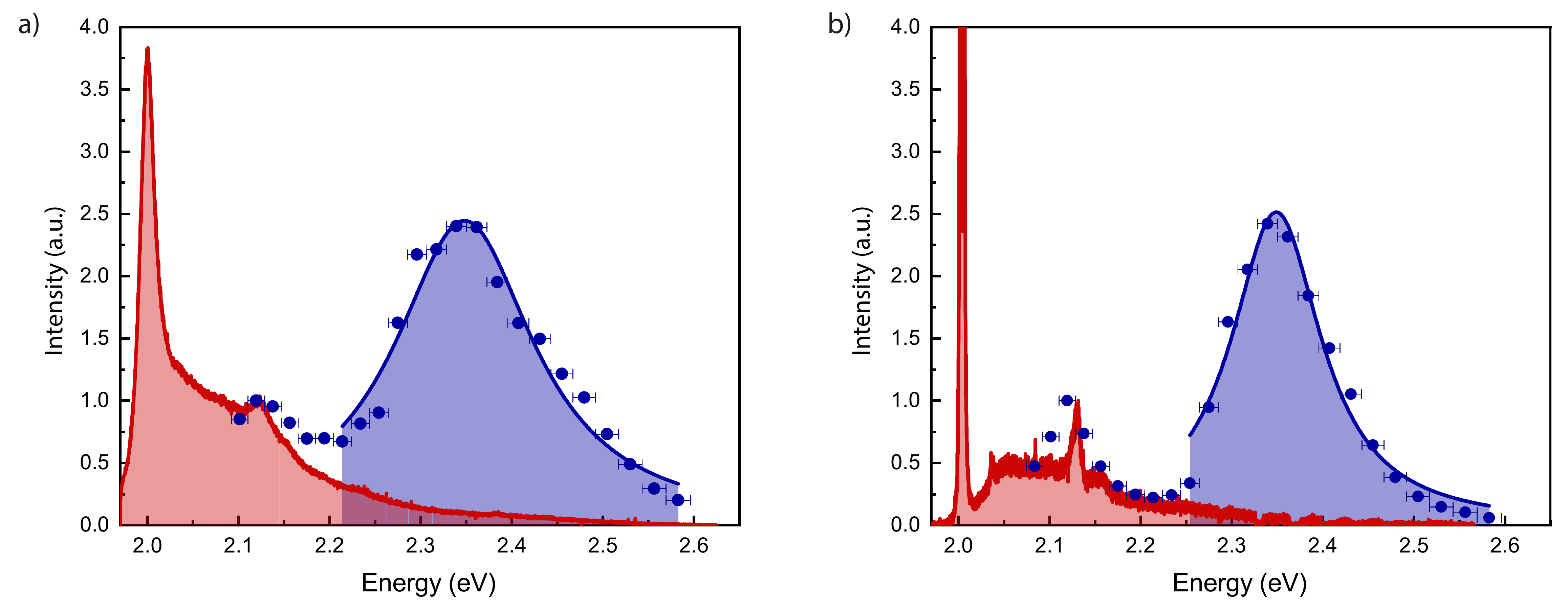}
  \caption{PLE measurements of an ensemble of $\text{SnV}^-$ centres. a) Room temperature: A broad resonance (blue shaded) lies centered around \SI{2.348(4)}{eV} with a width of \SI{190(12)}{meV} corresponding to an excitation from an $a_{2u}$ state in the valence band. A second peak is situated at \SI{2.12}{eV} coinciding with the most prominent sideband peak of the $\text{SnV}^-$. This is indicated by the mirror image (red shaded) of the PL spectrum with respect to the ZPL. b) \SI{1.6}{K}: The resonance remains at \SI{2.349(3)}{eV} but the width narrows down to \SI{120(8)}{meV}. This narrowing is due to a reduced homogeneous phonon broadening at low temperatures. The second peak is again coinciding with the most prominent sideband peak of the $\text{SnV}^-$.}
  \label{fig:MirrorImagePLE_RT_1p6}
\end{figure}

The evidence in literature \cite{HatanoTin, TchernijSnV, TrusheimTransLimSnV, RugarCharSnVPill} and in the present work strongly indicate a molecular model of the $\text{SnV}^-$ centre as inversion symmetric $\text{D}_{3\text{d}}$ point group. It thus should feature the same electronic structure as the well investigated $\text{SiV}^-$ and $\text{GeV}^-$ centres \cite{ThieringGroupIV}. The existence of $\text{E}_\text{g}$ and $\text{E}_\text{u}$ ground and excited states are confirmed by PL experiments (see Fig. \ref{fig:WaterfallS2}) and Zeeman splittings in external B-fields \cite{TrusheimTransLimSnV, RugarCharSnVPill}. In order to further explore the $\text{SnV}^-$ electronic level structure, we here perform PLE measurements on an ensemble in sample NI58. The excitation radiation is provided by a white light laser (SuperK Extreme EXW-12, NKT photonics), which we scan from \SI{480}{nm} up to \SI{595}{nm} while keeping its bandwidth fixed to \SI{5}{nm} and maintaining constant power. At room temperature, a spectrum is taken for every excitation wavelength and the counts constituting the ZPL of the $\text{SnV}^-$ are evaluated by fitting the spectra with Voigt profiles. Fig. \ref{fig:MirrorImagePLE_RT_1p6} a) shows a broad Lorentzian shaped resonance centered around \SI{2.348(4)}{eV} with a width of \SI{190(12)}{meV}. We attribute this resonance to a higher lying excited state with $A_{2u}$ symmetry resulting from a broad $a_{2u}$ band. This band is created by an $a_{2u}$ defect level situated in the valence band and therefore being broadened due to strong mixing with the diamond band. Similar results were found for the $\text{SiV}^-$ centre in diamond \cite{GaliSiVElecStrRelProp, HaeusslerPLESiVGeV}. The width of the $A_{2u}$ resonance (\SI{190}{meV}) being narrower than the corresponding state of the $\text{SiV}^-$ centre (\SI{400}{meV}) indicates a position of the $a_{2u}$ orbital closer to the valence band edge in agreement to the predictions of \cite{ThieringGroupIV}. At \SI{2.12}{eV} another peak is observable which corresponds to the most prominent sideband peak of the $\text{SnV}^-$ at \SI{122}{meV}. This is indicated in Fig. \ref{fig:MirrorImagePLE_RT_1p6} a) by overlapping the PLE measurement with the mirror image of the $\text{SnV}^-$ PL spectrum with respect to the ZPL position. The peak at \SI{2.12}{eV} is a result of addressing the corresponding higher lying vibrational ladder of the excited state and, furthermore, stating that the ground and excited states vibronic structure are of similar nature.
\newline
The measurement is repeated at \SI{2}{K}, detecting the $\text{SnV}^-$ ZPL by filtering with a \SI{620/10}{nm} bandpass filter. As shown in Fig. \ref{fig:MirrorImagePLE_RT_1p6} b), the absorption peak remains centered at \SI{2.349(3)}{eV} but with a decreased width of \SI{120(8)}{meV}. This narrowing of the transition is due to a reduced homogeneous phonon broadening at low temperatures. We thus conclude that the actual width of the transition is on the order of \SI{120}{meV}. In analogy to the case at room temperature, a second peak occurs also at \SI{2}{K}, again coinciding with the most prominent sideband peak present in the pure PL spectrum, thus supporting the explanation for the room temperature measurement.

\section{Summary and Conclusion}
In this work, we have investigated the spectroscopic properties of the $\text{SnV}^-$ centre. We found that the main dipole of the $\text{SnV}^-$ centre is aligned along the $\langle 111\rangle$ direction of the diamond lattice, thereby indicating the high symmetry axis of the defect. The electron-phonon scattering characteristics of the centre were investigated by measurement of ZPL line shifts and broadenings as function of temperature, by comparing the simulated (see \cite{ThieringGroupIV}) and the measured PSB as well as by the temperature dependence of the Debye-Waller factor of a single $\text{SnV}^-$ centre. From the results we were able to show that the electron-phonon interaction is dominated by phonons of $a_{1g}$ symmetry and that Debye-Waller factors can reach up to \SI{60}{\percent} at \SI{10}{K}. This value for the Debye-Waller factor is slightly smaller as in the case of the $\text{SiV}^-$ centre, but still more than a magnitude larger than for the $\text{NV}^-$ centre. By performing PLE spectroscopy over a large wavelength range, we revealed a higher lying excited state at \SI{2.35}{eV} resulting from an $a_{2u}$ defect level lying in the valence band and strongly coupling to the diamond modes. In the same measurement we found that the vibrational ladder of the ground state strongly resembles the one of the excited state indicating similar electronic wave functions.
Finally, single centres with truly Fourier limited linewidths were found with their usefulness being limited by a dominant ionisation process resulting in termination of the fluorescence when approaching the resonance maximum of the fine structure transitions and non negligible spectral diffusion. This can be overcome in future experiments by carefully controlling the Fermi level of the surrounding diamond host, for example by co-implantation of electron donors. Stabilising the negative charge state of the $\text{SnV}^-$ centre would further allow for measuring the ground state spin coherence time and comparing it to the theoretical expectation.

\section{Acknowledgement}
We thank Elke Neu and Richard Nelz for many helpful discussions. 
This  research  has  been  partially  funded  by  the  European Quantum Technology Flagship Horizon 2020 (H2020-EU1.2.3/2014-2020) under Grant No.  820394 (ASTERIQS).
This project has further received partial funding from the EMPIR programme co-financed by the Participating States and from the European Union’s Horizon 2020 research and innovation programme under Grant No. 17FUN06 (SIQUST).
This work was supported by JST-PRESTO (Grant No. JPMJPR16P2).
A.G. acknowledges the National Excellence Program of Quantum-Coherent Materials Project (Hungarian NKFIH grant no. KKP129866), the EU QuantERA Q-Magine Project (grant no. 127889), and the National Quantum Technology Program (grant No. 2017-1.2.1-NKP-2017-00001).
\newpage

\section{References}

\bibliography{BibtexPaper}

\providecommand{\newblock}{}
\begin{thebibliography}{10}
\expandafter\ifx\csname url\endcsname\relax
  \def\url#1{{\tt #1}}\fi
\expandafter\ifx\csname urlprefix\endcsname\relax\def\urlprefix{URL }\fi
\providecommand{\eprint}[2][]{\url{#2}}

\bibitem{AtatureMatPlatRev}
{Atat{\"u}re} M, {Englund} D, {Vamivakas} N, {Lee} S~Y and {Wrachtrup} J 2018
  {\em Nature Reviews Materials\/} {\bf 3} 38--51

\bibitem{AwschalomQuTechSSSRev}
Awschalom D~D, Hanson R, Wrachtrup J and Zhou B~B 2018 {\em Nature Photonics\/}
  {\bf 12} 516--527 ISSN 1749-4893
  \urlprefix\url{https://doi.org/10.1038/s41566-018-0232-2}

\bibitem{Bar-GillNVCohTime}
Bar-Gill N, Pham L~M, Jarmola A, Budker D and Walsworth R~L 2013 {\em Nature
  Communications\/} {\bf 4} 1743 EP -- article
  \urlprefix\url{https://doi.org/10.1038/ncomms2771}

\bibitem{AbobeihOneSecNV}
Abobeih M~H, Cramer J, Bakker M~A, Kalb N, Markham M, Twitchen D~J and Taminiau
  T~H 2018 {\em Nature Communications\/} {\bf 9} 2552 ISSN 2041-1723
  \urlprefix\url{https://doi.org/10.1038/s41467-018-04916-z}

\bibitem{BeckerMilliK}
Becker J~N, Pingault B, Gro\ss{} D, G\"undo\ifmmode~\breve{g}\else \u{g}\fi{}an
  M, Kukharchyk N, Markham M, Edmonds A, Atat\"ure M, Bushev P and Becher C
  2018 {\em Phys. Rev. Lett.\/} {\bf 120}(5) 053603
  \urlprefix\url{https://link.aps.org/doi/10.1103/PhysRevLett.120.053603}

\bibitem{SukachevQMSiV}
Sukachev D~D, Sipahigil A, Nguyen C~T, Bhaskar M~K, Evans R~E, Jelezko F and
  Lukin M~D 2017 {\em Phys. Rev. Lett.\/} {\bf 119}(22) 223602
  \urlprefix\url{https://link.aps.org/doi/10.1103/PhysRevLett.119.223602}

\bibitem{SipahigilSPSwitch}
Sipahigil A, Evans R~E, Sukachev D~D, Burek M~J, Borregaard J, Bhaskar M~K,
  Nguyen C~T, Pacheco J~L, Atikian H~A, Meuwly C, Camacho R~M, Jelezko F,
  Bielejec E, Park H, Lon{\v c}ar M and Lukin M~D 2016 {\em Science\/} {\bf
  354} 847--850 ISSN 0036-8075 (\textit{Preprint}
  \eprint{https://science.sciencemag.org/content/354/6314/847.full.pdf})
  \urlprefix\url{https://science.sciencemag.org/content/354/6314/847}

\bibitem{WeinzetlCohContWaveMix}
Weinzetl C, G\"orlitz J, Becker J~N, Walmsley I~A, Poem E, Nunn J and Becher C
  2019 {\em Phys. Rev. Lett.\/} {\bf 122}(6) 063601
  \urlprefix\url{https://link.aps.org/doi/10.1103/PhysRevLett.122.063601}

\bibitem{SipahigilQuIntRemNV}
Sipahigil A, Goldman M~L, Togan E, Chu Y, Markham M, Twitchen D~J, Zibrov A~S,
  Kubanek A and Lukin M~D 2012 {\em Phys. Rev. Lett.\/} {\bf 108}(14) 143601
  \urlprefix\url{https://link.aps.org/doi/10.1103/PhysRevLett.108.143601}

\bibitem{HensenLoopholeBell}
Hensen B, Bernien H, Dr{\'e}au A~E, Reiserer A, Kalb N, Blok M~S, Ruitenberg J,
  Vermeulen R~F~L, Schouten R~N, Abell{\'a}n C, Amaya W, Pruneri V, Mitchell
  M~W, Markham M, Twitchen D~J, Elkouss D, Wehner S, Taminiau T~H and Hanson R
  2015 {\em Nature\/} {\bf 526} 682 EP --
  \urlprefix\url{https://doi.org/10.1038/nature15759}

\bibitem{NeuSPSiVNano}
Neu E, Steinmetz D, Riedrich-Möller J, Gsell S, Fischer M, Schreck M and
  Becher C 2011 {\em New Journal of Physics\/} {\bf 13} 025012
  \urlprefix\url{https://doi.org/10.1088%2F1367-2630%2F13%2F2%2F025012}

\bibitem{HeppElecStr}
Hepp C, M\"uller T, Waselowski V, Becker J~N, Pingault B, Sternschulte H,
  Steinm\"uller-Nethl D, Gali A, Maze J~R, Atat\"ure M and Becher C 2014 {\em
  Phys. Rev. Lett.\/} {\bf 112}(3) 036405
  \urlprefix\url{https://link.aps.org/doi/10.1103/PhysRevLett.112.036405}

\bibitem{BeckerUltrafastAllOpt}
Becker J~N, G{\"o}rlitz J, Arend C, Markham M and Becher C 2016 {\em Nature
  Communications\/} {\bf 7} 13512 EP -- article
  \urlprefix\url{https://doi.org/10.1038/ncomms13512}

\bibitem{FuSpecDiffNV}
Fu K~M~C, Santori C, Barclay P~E and Beausoleil R~G 2010 {\em Applied Physics
  Letters\/} {\bf 96} 121907 ISSN 0003-6951
  \urlprefix\url{https://doi.org/10.1063/1.3364135}

\bibitem{WoltersUltrafastSpecDiffNV}
Wolters J, Sadzak N, Schell A~W, Schr\"oder T and Benson O 2013 {\em Phys. Rev.
  Lett.\/} {\bf 110}(2) 027401
  \urlprefix\url{https://link.aps.org/doi/10.1103/PhysRevLett.110.027401}

\bibitem{HatanoTin}
Iwasaki T, Miyamoto Y, Taniguchi T, Siyushev P, Metsch M~H, Jelezko F and
  Hatano M 2017 {\em Phys. Rev. Lett.\/} {\bf 119}(25) 253601
  \urlprefix\url{https://link.aps.org/doi/10.1103/PhysRevLett.119.253601}

\bibitem{TchernijSnV}
Tchernij S~D, Herzig T, Forneris J, Küpper J, Pezzagna S, Traina P, Moreva E,
  Degiovanni I~P, Brida G, Skukan N, Genovese M, Jakšić M, Meijer J and
  Olivero P 2017 {\em ACS Photonics\/} {\bf 4} 2580--2586 (\textit{Preprint}
  \eprint{https://doi.org/10.1021/acsphotonics.7b00904})
  \urlprefix\url{https://doi.org/10.1021/acsphotonics.7b00904}

\bibitem{ThieringGroupIV}
Thiering G and Gali A 2018 {\em Phys. Rev. X\/} {\bf 8}(2) 021063
  \urlprefix\url{https://link.aps.org/doi/10.1103/PhysRevX.8.021063}

\bibitem{JahnkeElecPhonSiV}
Jahnke K~D, Sipahigil A, Binder J~M, Doherty M~W, Metsch M, Rogers L~J, Manson
  N~B, Lukin M~D and Jelezko F 2015 {\em New Journal of Physics\/} {\bf 17}
  043011 \urlprefix\url{https://doi.org/10.1088%2F1367-2630%2F17%2F4%2F043011}

\bibitem{PingaultCohDarkStates}
Pingault B, Becker J~N, Schulte C~H~H, Arend C, Hepp C, Godde T, Tartakovskii
  A~I, Markham M, Becher C and Atat\"ure M 2014 {\em Phys. Rev. Lett.\/} {\bf
  113}(26) 263601
  \urlprefix\url{https://link.aps.org/doi/10.1103/PhysRevLett.113.263601}

\bibitem{RogersCohPrepSiV}
Rogers L~J, Jahnke K~D, Metsch M~H, Sipahigil A, Binder J~M, Teraji T, Sumiya
  H, Isoya J, Lukin M~D, Hemmer P and Jelezko F 2014 {\em Phys. Rev. Lett.\/}
  {\bf 113}(26) 263602
  \urlprefix\url{https://link.aps.org/doi/10.1103/PhysRevLett.113.263602}

\bibitem{InamTrackEmRate}
Inam F~A, Edmonds A~M, Steel M~J and Castelletto S 2013 {\em Applied Physics
  Letters\/} {\bf 102} 253109 (\textit{Preprint}
  \eprint{https://doi.org/10.1063/1.4812711})
  \urlprefix\url{https://doi.org/10.1063/1.4812711}

\bibitem{LesikPerfPrefOrientNV}
Lesik M, Tetienne J~P, Tallaire A, Achard J, Mille V, Gicquel A, Roch J~F and
  Jacques V 2014 {\em Applied Physics Letters\/} {\bf 104} 113107
  (\textit{Preprint} \eprint{https://doi.org/10.1063/1.4869103})
  \urlprefix\url{https://doi.org/10.1063/1.4869103}

\bibitem{RugarCharSnVPill}
Rugar A~E, Dory C, Sun S and Vu\ifmmode \check{c}\else
  \v{c}\fi{}kovi\ifmmode~\acute{c}\else \'{c}\fi{} J 2019 {\em Phys. Rev. B\/}
  {\bf 99}(20) 205417
  \urlprefix\url{https://link.aps.org/doi/10.1103/PhysRevB.99.205417}

\bibitem{TrusheimTransLimSnV}
Trusheim M~E, Pingault B, Wan N~H, Gündo\u{g}an M, De~Santis L, Chen K~C,
  Walsh M, Rose J~J, Becker J~N, Lienhard B, Bersin E, Malladi G, Lyzwa D,
  Bakhru H, Walmsley I, Atatüre M and Englund D 2018 {\em arXiv\/}  1811.07777

\bibitem{MohtashamiSuitNVNDSpontEmExp}
Mohtashami A and Koenderink A~F 2013 {\em New Journal of Physics\/} {\bf 15}
  043017 \urlprefix\url{https://doi.org/10.1088%2F1367-2630%2F15%2F4%2F043017}

\bibitem{ArendSpecHole}
Arend C, Becker J~N, Sternschulte H, Steinm\"uller-Nethl D and Becher C 2016
  {\em Phys. Rev. B\/} {\bf 94}(4) 045203
  \urlprefix\url{https://link.aps.org/doi/10.1103/PhysRevB.94.045203}

\bibitem{AlkathaniSnVThermo}
Alkahtani M, Cojocaru I, Liu X, Herzig T, Meijer J, Küpper J, Lühmann T,
  Akimov A~V and Hemmer P~R 2018 {\em Applied Physics Letters\/} {\bf 112}
  241902 (\textit{Preprint} \eprint{https://doi.org/10.1063/1.5037053})
  \urlprefix\url{https://doi.org/10.1063/1.5037053}

\bibitem{HizhnyakovZPLElasticSprings}
Hizhnyakov V, Kaasik H and Sildos I 2002 {\em phys. stat. sol. (b)\/} {\bf 234}
  644--653
  \urlprefix\url{doi:10.1002/1521-3951(200211)234:2<644::AID-PSSB644>3.0.CO;2-E}

\bibitem{HizhnyakovTempDepZPLI}
Hizhnyakov V, Boltrushko V, Kaasik H and Sildos I 2003 {\em The Journal of
  Chemical Physics\/} {\bf 119} 6290--6295 (\textit{Preprint}
  \eprint{https://doi.org/10.1063/1.1603216})
  \urlprefix\url{https://doi.org/10.1063/1.1603216}

\bibitem{HizhnyakovTempDepZPLII}
Hizhnyakov V, Boltrushko V, Kaasik H and Sildos I 2004 {\em Journal of
  Luminescence\/} {\bf 107} 351 -- 358 ISSN 0022-2313 proceedings of the 8th
  International Meeting on Hole Burning, Single Molecule, and Related
  Spectroscopies: Science and Applications
  \urlprefix\url{http://www.sciencedirect.com/science/article/pii/S0022231303002035}

\bibitem{LindnerStrInhSiV}
Lindner S, Bommer A, Muzha A, Krueger A, Gines L, Mandal S, Williams O, Londero
  E, Gali A and Becher C 2018 {\em New Journal of Physics\/} {\bf 20} 115002
  ISSN 1367-2630 \urlprefix\url{http://dx.doi.org/10.1088/1367-2630/aae93f}

\bibitem{NeuFluoPol}
Neu E, Fischer M, Gsell S, Schreck M and Becher C 2011 {\em Phys. Rev. B\/}
  {\bf 84}(20) 205211
  \urlprefix\url{https://link.aps.org/doi/10.1103/PhysRevB.84.205211}

\bibitem{NguyenNanophotQuRegSiV}
Nguyen C~T, Sukachev D~D, Bhaskar M~K, Machielse B, Levonian D~S, Knall E~N,
  Stroganov P, Chia C, Burek M~J, Riedinger R, Park H, Lon{\v c}ar M and Lukin
  M~D 2019 {\em arXiv\/}  1907.13200v2

\bibitem{LonderoVibModesSiV}
Londero E, Thiering G, Razinkovas L, Gali A and Alkauskas A 2018 {\em Phys.
  Rev. B\/} {\bf 98}(3) 035306
  \urlprefix\url{https://link.aps.org/doi/10.1103/PhysRevB.98.035306}

\bibitem{DietrichIsoShiftSiV}
Dietrich A, Jahnke K~D, Binder J~M, Teraji T, Isoya J, Rogers L~J and Jelezko F
  2014 {\em New Journal of Physics\/} {\bf 16} 113019
  \urlprefix\url{https://doi.org/10.1088%2F1367-2630%2F16%2F11%2F113019}

\bibitem{EkimovIsoShiftGeV}
Ekimov E~A, Lyapin S~G, Boldyrev K~N, Kondrin M~V, Khmelnitskiy R, Gavva V~A,
  Kotereva T~V and Popova M~N 2015 {\em JETP Letters\/} {\bf 102} 701--706 ISSN
  1090-6487 \urlprefix\url{https://doi.org/10.1134/S0021364015230034}

\bibitem{brand}
Brand J, Weinzierl G and Friedrich J 1981 {\em J. Chem. Phys. Lett.\/} {\bf 84}
  197--200

\bibitem{Plakhotnik_OptThermNV}
Plakhotnik T, Doherty M~W, Cole J~H, Chapman R and Manson N~B 2014 {\em Nano
  Letters\/} {\bf 14} 4989--4996 ISSN 1530-6984
  \urlprefix\url{https://doi.org/10.1021/nl501841d}

\bibitem{GaliSiVElecStrRelProp}
Gali A and Maze J~R 2013 {\em Phys. Rev. B\/} {\bf 88}(23) 235205
  \urlprefix\url{https://link.aps.org/doi/10.1103/PhysRevB.88.235205}

\bibitem{HaeusslerPLESiVGeV}
Häu{\ss}ler S, Thiering G, Dietrich A, Waasem N, Teraji T, Isoya J, Iwasaki T,
  Hatano M, Jelezko F, Gali A and Kubanek A 2017 {\em New Journal of Physics\/}
  {\bf 19} 063036 \urlprefix\url{https://doi.org/10.1088%2F1367-2630%2Faa73e5}

\end{thebibliography}

\end{document}